\documentclass[10pt]{iopart}
\usepackage{graphicx,iopams,epsfig,amssymb}

\begin{document}

\title[Dynamics of DFB dye lasing...]{Dynamics of DFB dye lasing by polarization modulation:
simulations and experiment}

\author{Denis V Novitsky, Vasili M Katarkevich and Terlan Sh Efendiev}
\address{B. I. Stepanov Institute of Physics, National Academy of
Sciences of Belarus, Nezavisimosti Avenue 68, BY-220072 Minsk,
Belarus} \ead{dvnovitsky@gmail.com}

\begin{abstract}
Distributed feedback (DFB) dye lasing by polarization modulation is
studied theoretically on the basis of modified rate equations.
Numerical solution of these equations allows to obtain the generated
power dynamics and the dependence of laser energy on pump energy.
The results of calculations are in good qualitative agreement with
the experimental data.
\end{abstract}


\vspace{2pc}


\submitto{Laser Phys. Lett.}

\maketitle

\ioptwocol

\section{Introduction}

Distributed feedback (DFB) lasers are compact and easy-to-use
sources of the narrowband coherent radiation. In the DFB lasers, the
feedback is provided via backward Bragg light scattering from the
spatially periodic structure formed inside the active medium itself
and, hence, is distributed throughout its length. The idea of such
lasers has been put forward and implemented by the authors of Refs.
\cite{Kogelnik, Shank}. As applied to dye solutions as active media,
the spatially periodic structures can be either of stationary
\cite{Kogelnik} or dynamic (rapidly reversible) \cite{Shank}
character. It should be noted that the dynamic DFB dye lasers
exhibit the highest performance characteristics. A number of optical
schemes of such lasers were developed. Their advantages had been
demonstrated, such as a small spectral linewidth, wide tuning range,
high efficiency, possibility of the high repetition rate operation
without dye solution flow as well as simultaneous generation at two
or more wavelengths with independent spectral tuning, etc.
\cite{Chandra, Zlenko, Efendiev1974, Bakos, Efendiev1975,
Rubinov1977, Efendiev1977, Bor1978, Rubinov1985}. At the same time,
the most interesting and important feature of the dynamic DFB dye
lasers is their ability to produce picosecond pulses both at
picosecond and nanosecond excitation \cite{Zaporozhchenko, Bushuk,
Bor1979, Bor1980, Masilamani, Efendiev1985, Ermilov}. In the latter
case, the use of such lasers represents the simplest way for
obtaining single pulses of a few tens of picoseconds duration with
possibility of smooth wavelength tuning.

Due to their important merits, DFB dye lasers are still the subject
of steady attention and intensive study by scientists from different
research centers (see some recent publications \cite{Diao, Toffanin,
Chida, Vannahme, Goldenberg, Silva, Smirnova, Sakhno}).

In the dynamic DFB dye lasers, excitation of the active medium is
generally provided by two converging pump beams with vertical
orientation of the electric field vector (s-polarization). In this
case, spatially periodic modulation of the resultant pump field
intensity is provided along the excited dye region. Fundamentally
different type of modulation occurs when the pump beams are
orthogonally polarized (for example, the first one is s-polarized,
while the second one is p-polarized). Under such conditions, the
pump field intensity is uniform over the excited zone of the dye,
while a periodic change of the resultant excitation field
polarization takes place. Due to the anisotropy of light absorption
and emission by dye molecules, the above pump field polarization
modulation results in the formation of a transient gain dichroism
(i.e., polarization) grating inside the dye solution.

Dynamic DFB dye laser action by polarization modulation was
experimentally investigated in Refs. \cite{Lo, Ye, Wang, Chen} where
spectral, threshold and polarization characteristics of a DFB laser
pumped by the second harmonic from a nanosecond Nd:YAG laser were
reported. Unlike the works mentioned above, in our studies we used
the second harmonic generation from a subnanosecond diode-pumped
solid-state (DPSS) Nd:LSB microlaser to initiate the DFB lasing by
polarization modulation in the dye solution \cite{Katarkevich1,
Katarkevich2}. Under such condition, the DFB laser exhibited a
spectral linewidth of $\Delta \lambda_{0.5} < 0.008$ nm (full width
at half-maximum, FWHM) and the energy conversion efficiency as high
as $\eta_{\rm max} \sim 48 \%$. It should be noted for comparison
that, under identical pumping conditions, the DFB lasing by
intensity modulation revealed the similar value of  $\Delta
\lambda_{0.5}$ and $\eta_{\rm max} \sim 50 \%$. Thus, in terms of
the spectral linewidth and peak lasing efficiency, the DFB laser
based on the dynamic polarization grating is comparable with that
employing the "traditional" (i.e., intensity modulation) pumping
geometry. To the best of our knowledge, no similar results regarding
spectral line narrowness and efficiency of the DFB dye lasing by
polarization modulation have been reported ever before. At the same
time, our results seem to be of an undoubted interest and indicate
the potential of these devices. It is obvious that, for its
practical realization, a deeper insight into the operation of such
DFB dye lasers is required. The latter is hardly possible without
appropriate theoretical model capable to adequately predict their
performance. Since, as far as we know, all existing studies still
lack for any theoretical treatment of the properties of such laser
devices, here we attempt to fill this gap.

In this paper, we propose a simple theoretical model of DFB lasing
by polarization modulation and report on the numerical simulations
of its dynamics for the case of subnanosecond excitation. The
results obtained reveal that, at moderate pumping rates, a
multi-spike emission is provided by a DFB laser, while single
picosecond pulse generation takes place at the pump intensities not
far from the threshold. As a proof of consistency and usefulness of
the proposed model, the output characteristics of the Rhodamine 6G
DFB laser excited by two orthogonally polarized second harmonic
beams from a subnanosecond DPSS Nd:LSB microlaser are presented.

\section{\label{model}Theoretical model}

\begin{figure}[t!]
\centering \includegraphics[scale=1, clip=]{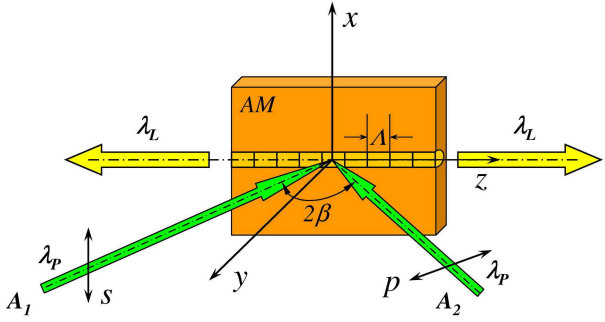}
\caption{\label{fig1} (Colour online) Schematic of the DFB laser
excitation by two crossing orthogonally polarized pumping beams. AM,
active medium; $\lambda_P$, pumping wavelength; $\lambda_L$, lasing
wavelength; $2\beta$, intersection angle of the pump beams inside
the medium; $\Lambda$, spatial modulation period; $A_1$ and $A_2$,
amplitudes of the pump waves.}
\end{figure}

The geometry of the model is shown in Fig. \ref{fig1}. The field of
the pump beam which forms DFB grating is a sum of two plane waves
crossing inside the medium at an angle $2 \beta$. We denote $\beta$
the angle between the pump wave vector and the normal to the surface
of the sample (plane $x - z$). We assume that the first pump wave is
polarized along $x$ axis (s-polarization), the second one is
polarized in the plane $y - z$ (p-polarization), and the wave
generated by the laser propagates along $z$ axis. Then, spatial
distribution of the total pump electric field can be written (after
time averaging) as follows,
\begin{eqnarray}
\vec{E} &=& \vec{e}_x A_1 \rme^{-\rmi k (y \cos \beta + z \sin
\beta)} + \nonumber \\
&+& (\vec{e}_y \sin \beta + \vec{e}_z \cos \beta) A_2 \rme^{-\rmi k
(y \cos \beta - z \sin \beta)}, \label{field}
\end{eqnarray}
where $A_1$ and $A_2$ are the amplitudes of the pump waves,
$\vec{e}_{x,y,z}$ are the unit vectors along the corresponding axes,
$k=2 \pi/\lambda_P$ is the wavenumber of the pump waves with the
wavelength $\lambda_P$. The field (\ref{field}) gives the periodic
change of the polarization state along $z$ axis, while its intensity
$|\vec{E}|^2$ does not depend on position.

Interaction of a dye molecule with the field is governed by the
value of $|\vec{d} \cdot \vec{E}|^2$, the direction of the molecular
dipole moment vector being defined as
\begin{eqnarray}
\vec{d} = |d| (\vec{e}_x \sin \theta \cos \varphi + \vec{e}_y \cos
\theta \sin \varphi + \vec{e}_z \cos \theta), \label{dipole}
\end{eqnarray}
where the angles of the spherical coordinate system $\theta$ (zenith
angle) and $\varphi$ (azimuth angle) are counted from $z$ and $x$
axes, correspondingly. Using Eqs. (\ref{field}) and (\ref{dipole}),
the excitation efficiency for the molecule with dipole moment
oriented along angles $\theta$ and $\varphi$ can be calculated by
\begin{eqnarray}
f_{\rm{exc}} (\theta, \varphi) &=& |\vec{d} \cdot \vec{E}|^2 /
|\vec{d}|^2 |\vec{E}|^2 = \nonumber \\ &=& A (\theta, \varphi) + B
(\theta, \varphi) \cos (2 \pi z/\Lambda). \label{excite}
\end{eqnarray}
Here $\Lambda=\pi / k \sin \beta = \lambda_P / 2 \sin \beta$ is the
grating period, and the auxiliary functions are
\begin{eqnarray}
A (\theta, \varphi) &=& I_1 \sin^2 \theta \cos^2 \varphi + I_2
\left( \sin^2 \beta \sin^2 \theta \sin^2 \varphi + \right. \nonumber
\\
&+& \left. \cos^2 \beta \cos^2 \theta + \frac{1}{2} \sin 2\beta \sin
2\theta \sin \varphi \right), \nonumber
\end{eqnarray}
\begin{eqnarray}
B (\theta, \varphi) = \sqrt{I_1 I_2} (\sin \beta \sin^2 \theta \sin
2\varphi + \cos \beta \sin 2\theta \cos \varphi), \nonumber
\end{eqnarray}
where $I_{1,2}=A^2_{1,2}/(A_1^2 + A_2^2)$ are the coefficients which
define the part of the entire pump power contained in both incident
waves.

Thus, for every molecule orientation, we have the excitation grating
which is the result of polarization modulation and is described by
Eq. (\ref{excite}). In order to calculate lasing dynamics, we use
the idea of Ref. \cite{Rubinov2000}. Let us consider molecules of
given orientation separately in maxima and minima of the excitation
grating, i.e. for two values of efficiency $f^\pm_{\rm{exc}}
(\theta, \varphi) = A (\theta, \varphi) \pm |B (\theta, \varphi)|$.
These two values should be substituted in the rate equations which
describe temporal dynamics of the excited molecules concentration
and density of generated photons. To obtain these equations, we use
the rate equations given in Refs. \cite{Bor1980, Bor1982} for DFB
laser action by intensity modulation and generalize them on the case
of polarization modulation. As a result, we have the following
equations for the densities $n_\pm (\theta, \varphi)$ of excited
molecules in maxima and minima of the excitation grating,
\begin{eqnarray}
\frac{{\rm d} n_\pm (\theta, \varphi)}{{\rm d} t} &=& 3 I_p(t)
\sigma_a f^\pm_{\rm{exc}} (\theta, \varphi) [N/4\pi - n_\pm (\theta,
\varphi)] - \nonumber \\
&-& \frac{n_\pm (\theta, \varphi)}{\tau} - \label{concmol} \\
&-& 3 \frac{c}{\eta} \sigma_e n_\pm (\theta, \varphi) (q_x \cos^2
\varphi + q_y \sin^2 \varphi) \sin^2 \theta,\nonumber
\end{eqnarray}
where $\sigma_a$ and $\sigma_e$ are the absorption and emission
cross-sections; $I_p(t)$ is the intensity profile of the pump pulse;
$q_x$ and $q_y$ are the average densities of photons polarized along
$x$ and $y$ axes, respectively; $N$ is the density of dye molecules;
$\tau$ is the life-time of excited state; $\eta$ is the refractive
index of the dye solution; $c$ is the vacuum speed of light; the
coefficient $3$ is the result of usage of the cross-sections
averaged over molecular orientations. The first term on the right
side of Eq. (\ref{concmol}) describes the process of molecular
excitation under pump influence, while the second and third terms
are responsible for spontaneous and stimulated decay of the excited
state, respectively.

In order to obtain the equations for $q_x$ and $q_y$, we need the
expressions for average gain coefficients for light polarized along
$x$ and $y$ axes (s- and p-polarizations). These values are defined
as
\begin{eqnarray}
k_{x, y}=(k_{x, y}^+ + k_{x, y}^-)/2, \label{coefampl}
\end{eqnarray}
where the gain coefficients in the maxima and minima of the
excitation grating can be calculated through
\begin{eqnarray}
k_x^\pm = 3 \sigma_e \int\limits_0^\pi \int\limits_0^{2\pi} n_\pm
(\theta, \varphi) \cos^2 \varphi \sin^3 \theta \,{\rm
d}\varphi\,{\rm d}\theta, \nonumber \\
k_y^\pm = 3 \sigma_e \int\limits_0^\pi \int\limits_0^{2\pi} n_\pm
(\theta, \varphi) \sin^2 \varphi \sin^3 \theta \,{\rm
d}\varphi\,{\rm d}\theta. \nonumber
\end{eqnarray}
Then, the equations for average densities of s- and p-polarized
photons are as follows,
\begin{eqnarray}
\frac{{\rm d} q_{x, y}}{{\rm d} t} = \frac{c}{\eta} k_{x, y} q_{x,
y} - \frac{q_{x, y}}{\tau^{x, y}_c} + \frac{\Omega}{\tau} \bar{n}.
\label{concphot}
\end{eqnarray}
Here, the photon life-time in the cavity $\tau^{x, y}_c=\eta L^3
\alpha_{x, y}^2/8 c \pi^2$ is proportional to squared amplitude of
the gain grating \cite{Bor1980} which, in our case, is equal to
$\alpha_{x, y}=(k_{x, y}^+ - k_{x, y}^-)/2$; $L$ is the length of
the DFB structure. The last term in Eq. (\ref{concphot}) describes
spontaneous emission \cite{Bor1982} with a coefficient $\Omega=b/\pi
N \sigma_a L^2 S$, where $b$ is the height of the excited volume,
$S$ is the spectral factor which determines the fraction of
spontaneous emission falling into the laser bandwidth. The average
density of excited molecules is calculated by
$\bar{n}=(<n_+>+<n_->)/2$ with $<n_\pm>=\int\limits_0^\pi
\int\limits_0^{2\pi} n_\pm (\theta, \varphi) \sin \theta \,{\rm
d}\varphi\,{\rm d}\theta$.

Finally, lasing power can be determined by the expression
\cite{Bor1980, Bor1982}
\begin{eqnarray}
P_{\rm out} = P_x(t) + P_y(t) = \frac{1}{2} \frac{h c}{\lambda_g} L
b a \left[ \frac{q_x (t)}{\tau^x_c (t)} + \frac{q_y (t)}{\tau^y_c
(t)} \right], \label{power}
\end{eqnarray}
where $h$ is the Planck constant, and the penetration depth $a$ of
pump beam into the dye solution can be estimated from
\begin{eqnarray}
\frac{1}{a} = 3 \sigma_a \frac{N}{4 \pi} \int\limits_0^\pi
\int\limits_0^{2\pi} A (\theta, \varphi) \sin \theta \,{\rm
d}\varphi\,{\rm d}\theta = \sigma_a N. \nonumber
\end{eqnarray}

Equations (\ref{concmol}) and (\ref{concphot}) are the main
relations of our model. Solving them numerically and substituting
this solution into Eq. (\ref{power}), we can describe dynamics of
power and the polarization properties of DFB lasing by polarization
modulation.

\section{\label{calc}Numerical results}

\begin{figure*}[t!]
\centering \includegraphics[scale=0.9, clip=]{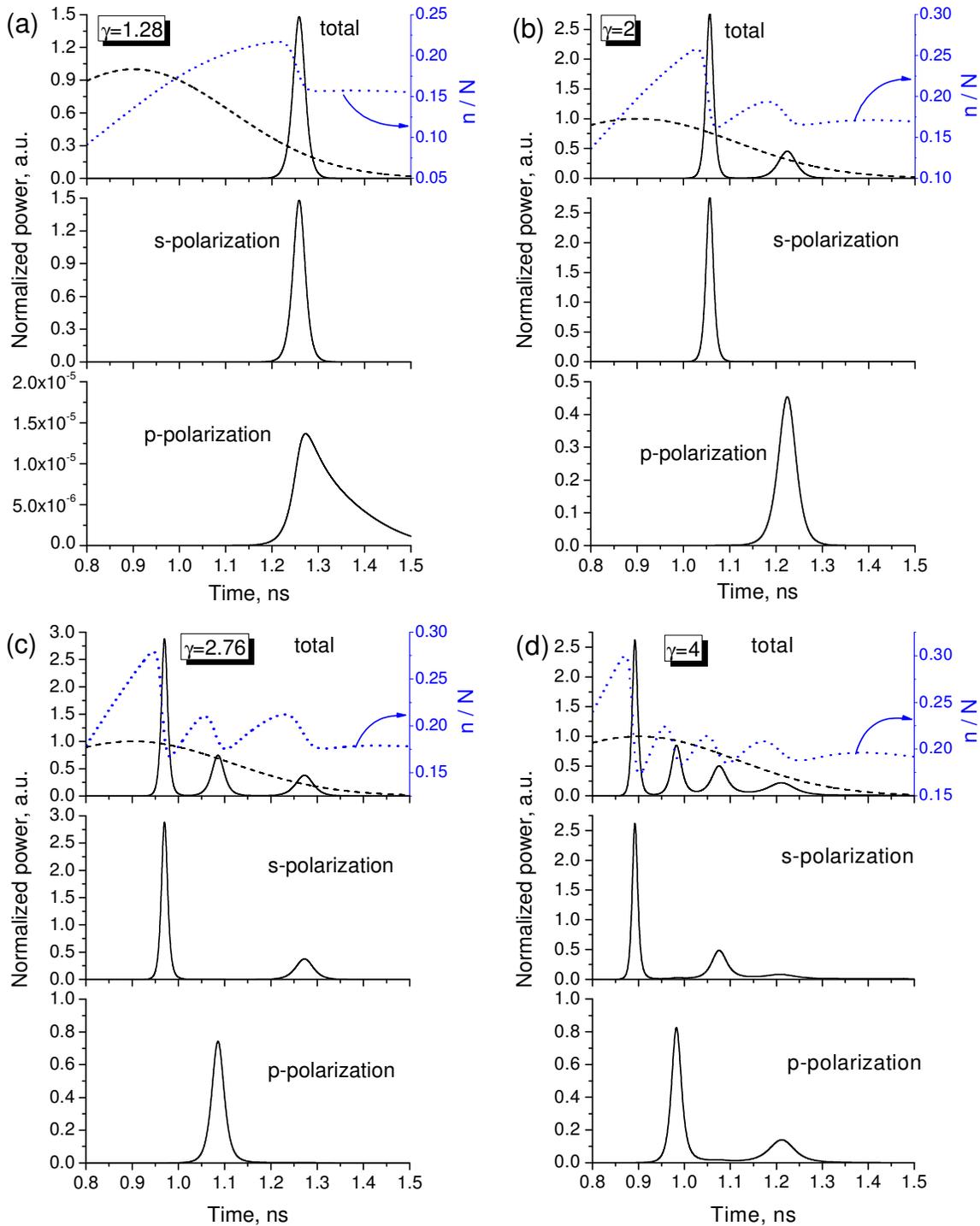}
\caption{\label{fig2} (Colour online) Results of calculation of the
generated pulses profiles (solid lines) and dynamics of average
density of excited molecules (dotted lines) at different pump
levels: (a) $\gamma=1.28$, (b) $2$, (c) $2.76$, (d) $4$. The density
of excited molecules is normalized by the full density of dye
molecules $N$. Dashed line shows the profile of the pump pulse.}
\end{figure*}

Let us consider some results of numerical solution of the rate
equations presented in the previous section. Calculations were
performed for the set of parameters which can be considered typical
for the dye-laser systems: pump wavelength $\lambda_P=532$ nm, laser
wavelength $\lambda_L=565$ nm, absorption and emission
cross-sections (Rhodamine 6G) $\sigma_a=3.8 \cdot 10^{-16}$ cm$^2$
and $\sigma_e=2.15 \cdot 10^{-16}$ cm$^2$, concentration of the dye
solution $C=0.243$ mM, refractive index of the solution at the laser
wavelength $\eta=1.36$, excited state life-time $\tau=4$ ns, the
length and the height of DFB structure $L=1$ cm and $b=0.01$ cm,
respectively, spectral factor $S=10^4$. The pulse of pump intensity
has the Gaussian envelope with FWHM $0.5$ ns. The ratio of
intensities of pump beams with s- and p-polarizations is $2:1$, i.e.
the intensity parameters are $I_1=0.67$ and $I_2=0.33$. The angle of
incidence for pump beams is $\beta=\arcsin
\lambda_P/\lambda_L=70.32^\circ$.

We start our analysis with the laser power dynamics calculated at
different pumping levels. We are interested rather in qualitative
relations than in absolute values of laser characteristics.
Therefore, we measure pump energy $W_{\rm p}$ with normalized
parameter $\gamma=W_{\rm p}/W_{\rm th}$, where $W_{\rm th}$ is the
threshold pump energy. This threshold is determined as the point of
abrupt rise of the output power. In absolute units, our calculations
give $W_{\rm th}=2.5$ $\mu$J for the parameters listed above.

According to our calculations, only s-polarized light is generated
at low excesses over pump threshold. In this case, laser radiation
forms a single pulse which, at threshold, appears at the trailing
edge of the pump pulse and arises more and more earlier as the pump
energy grows. The example of such s-polarized pulse is shown in Fig.
\ref{fig2}(a) at $\gamma=1.28$. This single-pulse lasing corresponds
to the sharp decrease of the average density $\bar{n}$ of excited
molecules presented in the same figure (dotted curve). It is seen
that, at first, the value $\bar{n}$ grows smoothly as a result of
pump and then, when it becomes larger than certain level (which, in
general, depends on pump energy), the density of excited molecules
drops releasing the stored energy in the form of laser radiation. As
to generation of p-polarized light under these conditions, it has
low intensity and cannot be described by the characteristic
bell-shaped envelope [see the lower panel in Fig. \ref{fig2}(a)].

When $\gamma \gtrsim 1.5$, generation of the second pulse starts.
This second pulse is p-polarized that is illustrated by the profiles
calculated at $\gamma=2$ [Fig. \ref{fig2}(b)]. In this case, the
dynamics of average density of excited molecules has two
well-pronounced peaks which correspond to both laser pulses. The
third pulse appears for $\gamma \gtrsim 2.25$ and is s-polarized as
the first one. The example of calculated power profiles and behavior
of $\bar{n}$ with three maxima are shown in Fig. \ref{fig2}(c) for
$\gamma=2.76$. Finally, when $\gamma \gtrsim 3$, the fourth pulse is
generated. As can be seen in Fig. \ref{fig2}(d), at $\gamma=4$, this
fourth pulse contains comparable quantities of both s- and
p-polarized radiation.

\begin{figure}[t!]
\centering \includegraphics[scale=0.85, clip=]{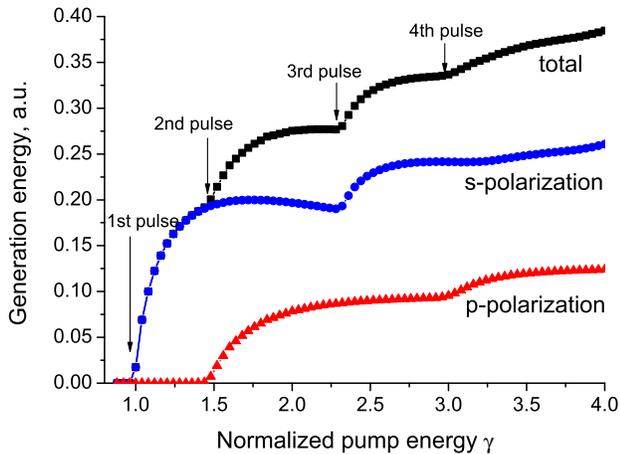}
\caption{\label{fig3} (Colour online) Laser output energy vs. pump
energy calculated with our model. The arrows indicate pump energies
where new pulse appears at the laser output.}
\end{figure}

After these examples, we should consider the energetic
characteristics of laser action described by our model. Figure
\ref{fig3} demonstrates the full output energy and the energy of s-
and p-polarized laser light taken separately as the functions of
pump energy. One can see that appearance of subsequent pulses is
characterized by inflection points in the curves depicted in Fig.
\ref{fig3}. Note that before such inflection point is reached, the
plateau occurs where generated energy changes very slightly with
pump energy. This property of the laser system considered seems to
be perspective to obtain linearly polarized single pulses with high
stability of output energy.

\section{\label{exper}Experimental results}

\begin{figure*}[t!]
\centering \includegraphics[scale=1, clip=]{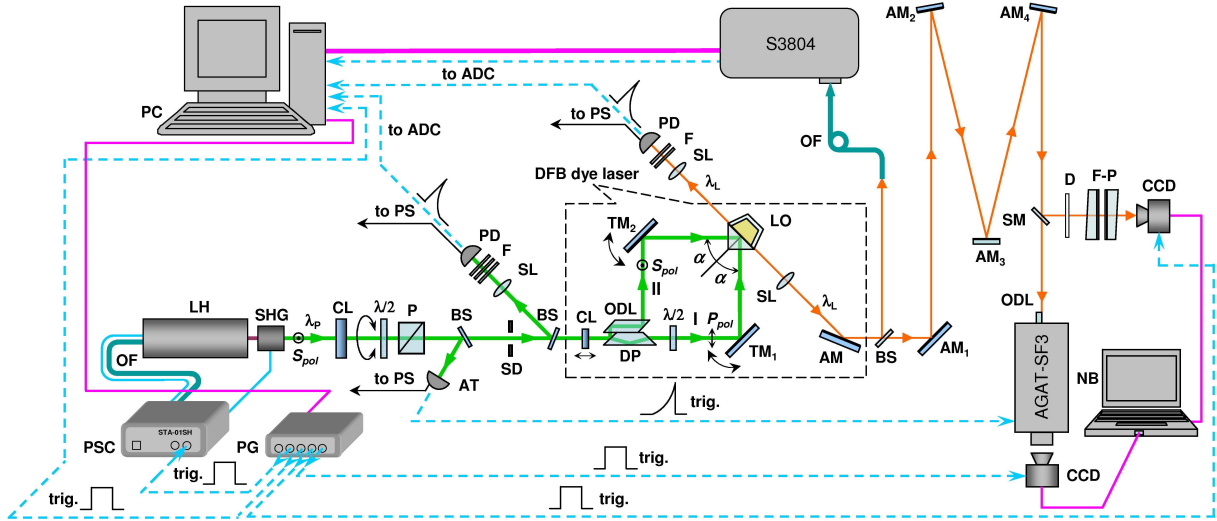}
\caption{\label{fig4} (Colour online) Schematic of the experimental
setup. PC, personal computer; NB, notebook; S3804, automated
diffraction grating spectrograph; OF, optical fiber; PSC, Nd:LSB
micro laser power supply and control unit; LH, Nd:LSB micro laser
head; SHG, second harmonic generator; PG, G-200P programmable TTL
pulse generator; Agat-SF3, streak camera; CCD, SDU-R205
charge-coupled device USB-TV camera; PD, FD-24K silicon photodiode;
AT, KT-342 hf avalanche transistor; ADC, analog-to-digital
converter; PS, power supply; F-P, IT51-30 Fabry-Perot
interferometer; D, diffuser; CL, cylindrical lens; SL, spherical
lens; F, optical filters; $\lambda/2$, half-wave plate; P,
dielectric polarizer; BS, beam splitter; SD, slit diaphragm; DP,
Dove prism beam splitter; ODL, optical delay line; TM, turning
dielectric mirror; LO, DFB laser oscillator; $\alpha$, half of the
pump beams I, II interference angle; $\lambda_P$, pumping
wavelength; $\lambda_L$, DFB lasing wavelength; AM, aluminum mirror;
SM, semi-transparent dielectric mirror.}
\end{figure*}

The output characteristics of a DFB dye laser based on the dynamic
polarization grating were investigated under subnanosecond
excitation. The optical scheme of a DFB laser and registering
apparatus were practically the same as in \cite{Efendiev2011,
Katarkevich3} (Fig. \ref{fig4}). A $0.26$ mM ethanol solution of
Rhodamine 6G was used as active medium. The vertically polarized
second harmonic generation ($\lambda=532$ nm; $\Delta
\lambda_{0.5}<0.003$ nm; beam quality factor $M^2<1.2$) from a DPSS
STA01SH-500 Nd:LSB microlaser (Standa Ltd., Lithuania) delivering
$\sim 0.5$ ns (FWHM) pulses with an energy of $E_P \leq 80$ $\mu$J
and a repetition rate as high as $f=500$ Hz was employed as a pump
source. The DFB laser oscillator represented a right-angle isosceles
K-8 glass prism hypotenuse side of which is in optical contact with
a dye in a cell with nonparallel windows. In such a scheme, the
$532$ nm pump beams are symmetrically coupled into the active medium
through the side faces of the prism thereby resulting in the
first-order DFB laser operation. Laser oscillations occur at a
wavelength $\lambda_L$ given by
\begin{eqnarray}
\lambda_L = \frac{n_s \lambda_P}{2 n_{\rm pr} \sin \alpha},
\end{eqnarray}
where where $n_s$ is the refractive index of the dye solution at the
lasing wavelength $\lambda_L$, $n_{\rm pr}$ is the refractive index
of the prism material at the pump wavelength $\lambda_P$; $\alpha$
is the incident angle of the pump beam at the prism-dye solution
interface (it differs from the angle $\beta$ inside the solution).
To obtain a pair of pump beams with s- and p-polarizations,
respectively, the electric field vector of one of the two microlaser
beams was turned by $90^\circ$ with the help of a half-wave plate.
The intensity of s-polarized beam was approximately two times higher
than that of p-polarized one. The maximum total pump beams energy
falling onto the surface of the dye solution in a cell did not
exceed $E_P=35$ $\mu$J.

To provide optimal pumping geometry, the output beam from a
microlaser with an initial diameter of about $0.02$ cm and a
full-angle divergence of $\sim 10$ mrad was collimated in the
vertical plane by a cylindrical lens (focus length $18$ cm), while
the distance between the microlaser head and the DFB laser input
window was set to $\sim 99$ cm. Upon such conditions, the excited
zone of the dye in a cell represented a narrow horizontal stripe
with a $\sim 0.9$ to $1.2$ cm length and $\sim 0.01$ cm height,
depending on the intersection angle $2 \alpha$ of the pump beams.

During experiments, the pump input and the DFB laser output energies
were measured simultaneously with calibrated FD-24K photodiodes and
an ADC-2OM/10-2 two-channel analog-to-digital converter (ADC). To
adjust the pump power in a continuous manner, a rotatable half-wave
plate in combination with a dielectric polarizer was employed. A
fiber-coupled S3804 automated diffraction grating spectrograph
(spectral resolution up to $\sim 0.08$ nm) and an IT51-30
Fabry-Perot interferometer were used for coarse and fine spectral
measurements, respectively. The transient behavior of the pump and
DFB laser pulses was studied using an Agat-SF3 ultrafast streak
camera (time resolution up to $\sim 2$ ps). Both the interferograms
and streak camera images were captured by means of a SDU-R205
charge-coupled device (CCD) USB-TV camera with subsequent computer
acquisition and processing of the obtained data.

Under mentioned above pumping conditions, a DFB laser exhibited
smooth wavelength tunability within $549-592$ nm upon a spectral
linewidth of $\Delta \lambda_{0.5}<0.008$ nm and an
optical-to-optical energy conversion efficiency up to $\eta_{\rm
max} \sim 60 \%$. Dependence of the output characteristics of a DFB
laser on the experimental pumping conditions was investigated in
details at $\lambda_L = 565$ nm, which falls into the dye
amplification band maximum. In this case, the length of the pumped
region of active medium was $L_{\rm DFB} \approx 1$ cm.

First of all, the influence of the pump pulse intensity on the
temporal course of the DFB laser output was studied in details. It
was found that, generally, when pumped well above threshold, the DFB
laser produces a multi-spike emission. Both an overall length and a
number of spikes in the pulse train depend on the pumping level
$\gamma$ and tend to decrease with decreasing the $\gamma$ value. At
pumping levels not very far from threshold single ultrashort pulses
of a few tens of picoseconds duration are generated.

\begin{figure}[t!]
\centering \includegraphics[scale=1, clip=]{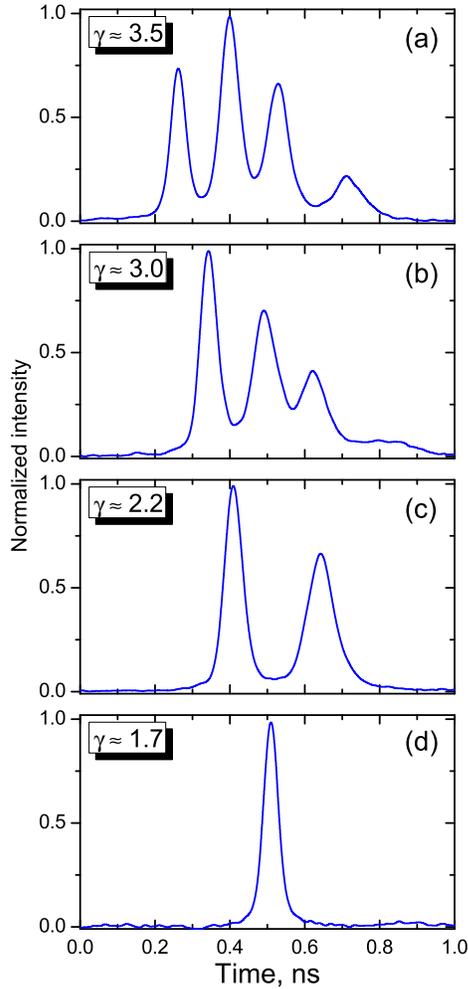}
\caption{\label{fig5} (Colour online) Streak camera traces of the
DFB dye laser pulses taken at different pumping levels $\gamma$.}
\end{figure}

As an example, Fig. \ref{fig5} reveals streak camera traces of the
DFB laser pulses registered at different pumping levels
corresponding to the next pulse appearing in the laser output.
Notice that, unlike Fig. \ref{fig2}, the zero position of the time
scale is of an arbitrary character and does not reflects the real
build-up dynamic of the DFB lasing. Should also be noted that, due
to relatively low streak camera scan rate ($v_{\rm sc} \sim 1$
ns/cm) used during these measurements, both the individual pulses in
the pulse trains and their durations are not fully resolved.

It is seen from Fig. \ref{fig5}(a) that at a pumping level of
$\gamma \approx 3.5$ a DFB laser produces a train of four pulses
with an overall width of $\sim 900$ ps. When the pump power is
lowered to $\gamma \approx 3.0$ [Fig. \ref{fig5}(b)] and $2.2$ [Fig.
\ref{fig5}(c)], the number of emitted pulses reduces to three and
two, respectively, while the overall pulse train lengths shorten to
$\sim 700-800$ ps. At pumping levels $1<\gamma<1.7$ stable
generation of single picosecond pulses was observed [Fig.
\ref{fig5}(d)]. The single pulse duration was rather sensitive to
the pumping level and progressively broadened with decreasing the
$\gamma$ value. The shortest pulses were obtained near threshold of
the second pulse ($\gamma \approx 1.7$). In this case, their
duration was measured to be about $\tau_{0.5} \approx 37$ ps (FWHM).
Notice that before performing above measurements streak camera was
set to operate with a scan rate of $v_{\rm sc} \sim 0.25$ ns/cm thus
providing an approximately four-fold increase in the time
resolution. Under such conditions, the spectral linewidth of the DFB
laser emission amounted to $\Delta \lambda_{0.5} \approx 0.0075$ nm
while its energy reached $E_L \sim 0.13$ $\mu$J. Based on these
data, the time-bandwidth product $\nu_{0.5} \tau_{0.5}$ and peak
power $P_L$ of a single pulse are estimated to be $0.3$ and $3.5$
kW, respectively. The obtained value of $\nu_{0.5} \tau_{0.5}$ is
the evidence of the transform-limited character of single picosecond
pulses generated by a DFB dye laser.

\begin{figure}[t!]
\centering \includegraphics[scale=0.9, clip=]{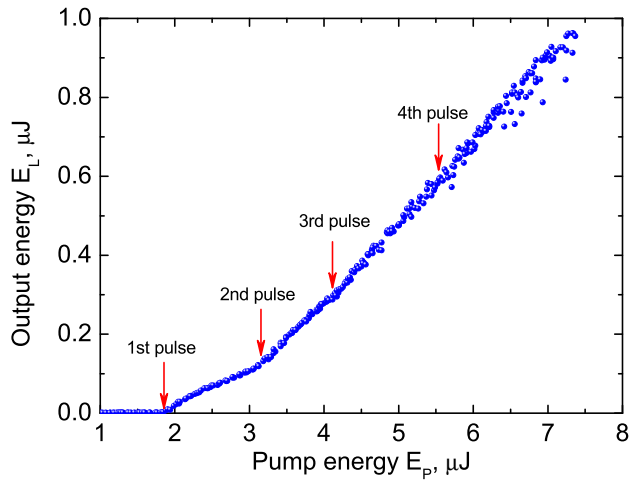}
\caption{\label{fig6} (Colour online) DFB dye laser output energy as
a function of pump energy.}
\end{figure}

Figure \ref{fig6} shows the dependence of the DFB laser output
energy $E_L$ on the pump energy $E_P$ measured for the range of
pumping levels $1<\gamma \leq 4$ (i.e., close to that of Fig.
\ref{fig5}). It is seen that $E_L$ gradually grows with increasing
the pump intensity. Along with this, at least three alterations in
slope may be noticed corresponding to the inflection points in Fig.
\ref{fig3}. The first one is observed around pump energy of $\sim
1.85$ $\mu$J representing the lasing threshold, while the second and
third ones correspond to the $E_P \approx 3.15$ and $4.1$ $\mu$J,
respectively. Based on the two latter values of pump energies, we
find that the appropriate pumping levels $\gamma$ are accordingly
$\sim 1.7$ and $\sim 2.2$. From streak camera measurements presented
in Fig. \ref{fig5}(d) and (c) follows that the specified pumping
level of $\gamma \approx 1.7$ and $2.2$ are indicative of the second
and third picosecond pulses appearing in the DFB laser output.

The above features regarding the course of the input-output
characteristic of a DFB laser operating not far from threshold seem
to be of undoubted practical interest. Indeed, based solely on the
energy measurement, it becomes possible without using a high-cost
streak camera to precisely determine the range of pump intensities
within which single picosecond pulse generation is provided by a DFB
dye laser.

\section{\label{concl}Conclusion}

In conclusion, we have proposed the theoretical model of DFB dye
laser action by polarization modulation. This fundamentally simple
model based on consideration of values averaged over excitation
grating allows to describe lasing dynamics, energetic and
polarization characteristics of such lasers. The mathematical basis
of the model is the modified rate equations for the density of
excited molecules and the concentration of laser photons. Comparison
of our calculation results with experimental data shows good
qualitative agreement between theory and experiment. In particular,
the model allows to describe the multi-spike lasing and
approximately predict the threshold values for successive pulse
generation. Polarization characteristics of such DFB lasers will be
studied in detail in a separate work.

As far as we know, we have performed the first realistic description
of lasing dynamics of DFB dye laser based on polarization
modulation. However, to obtain not only qualitative, but also
quantitative agreement between calculations and measurements, one
needs an advanced theory. In future, we plan to develop a
semiclassical treatment of such lasers which is expected to give
closer fit to the experimental data.

\section*{Acknowledgements}
The work was supported by Belarusian State Foundation for
Fundamental Research (Project F15-042).

\end{document}